\documentclass[runningheads]{llncs}
\usepackage{graphicx}
\usepackage{graphicx}
\usepackage{algorithm2e}
\usepackage{numprint}
\usepackage{url}
\usepackage{wrapfig}
\usepackage{adjustbox}

\usepackage{xcolor}
\usepackage{subfigure}
\usepackage{placeins}

\begin{document}

\def\Plus{\texttt{+}}
\npdecimalsign{.}
\nprounddigits{3}

\title{Med-NCA: Robust and Lightweight Segmentation with Neural Cellular Automata} 
%
%
\author{John Kalkhof\orcidID{0000-0001-7316-1903} \and 
Camila Gonz\'{a}lez\orcidID{0000-0002-4510-7309} \and 
Anirban Mukhopadhyay\orcidID{0000-0003-0669-4018}} 
\institute{Darmstadt University of Technology, Karolinenplatz 5, 64289 Darmstadt, Germany}
\authorrunning{J. Kalkhof et al.}
\titlerunning{Med-NCA}
%
%
\maketitle           

\let\thefootnote\relax\footnotetext{DOI will follow}

\begin{abstract}

Access to the proper infrastructure is critical when performing medical image segmentation with Deep Learning. This requirement makes it difficult to run state-of-the-art segmentation models in resource-constrained scenarios like primary care facilities in rural areas and during crises. The recently emerging field of Neural Cellular Automata (NCA) has shown that locally interacting \emph{one-cell} models can achieve competitive results in tasks such as image generation or segmentations in low-resolution inputs. However, they are constrained by high VRAM requirements and the difficulty of reaching convergence for high-resolution images. To counteract these limitations we propose Med-NCA, an end-to-end NCA training pipeline for high-resolution image segmentation. Our method follows a two-step process. Global knowledge is first communicated between cells across the downscaled image. Following that, patch-based segmentation is performed. Our proposed Med-NCA outperforms the classic UNet by 2\% and 3\% Dice for hippocampus and prostate segmentation, respectively, while also being \textbf{500 times smaller}. We also show that Med-NCA is by design invariant with respect to image scale, shape and translation, experiencing only slight performance degradation even with strong shifts; and is robust against MRI acquisition artefacts. Med-NCA enables high-resolution medical image segmentation even on a Raspberry Pi B\Plus, arguably the smallest device able to run PyTorch and that can be powered by a standard power bank.


\end{abstract}

\section{Introduction}
State-of-the-art medical image segmentation is dominated by UNet-style architectures \cite{UNet}, which still perform at the top of most grand challenges in its various forms \cite{nnUnet}. This trend of task-specific optimizations of UNet-style models is usually accompanied by diminishing returns regarding model size versus performance. The increase in model complexity raises serious concerns that machine learning cannot be leveraged in resource-constrained environments \cite{resourceConstrained_overview}. In settings such as primary care facilities in rural areas, only minimal computing infrastructure is available \cite{boppart2014point}, so it is challenging to deploy models requiring large GPUs. UNet-style models are also particularly susceptible to the \emph{domain shift} problem \cite{gonzalez2022distance} and have difficulty generalising to other input resolutions. To mitigate this, shifts are often brute-forced into the training pipeline by adding augmentations like translation or simulated acquisition artefacts \cite{chlap2021review}. 



\begin{figure}[htbp]
 \centering
  {\includegraphics[width=.85\linewidth]{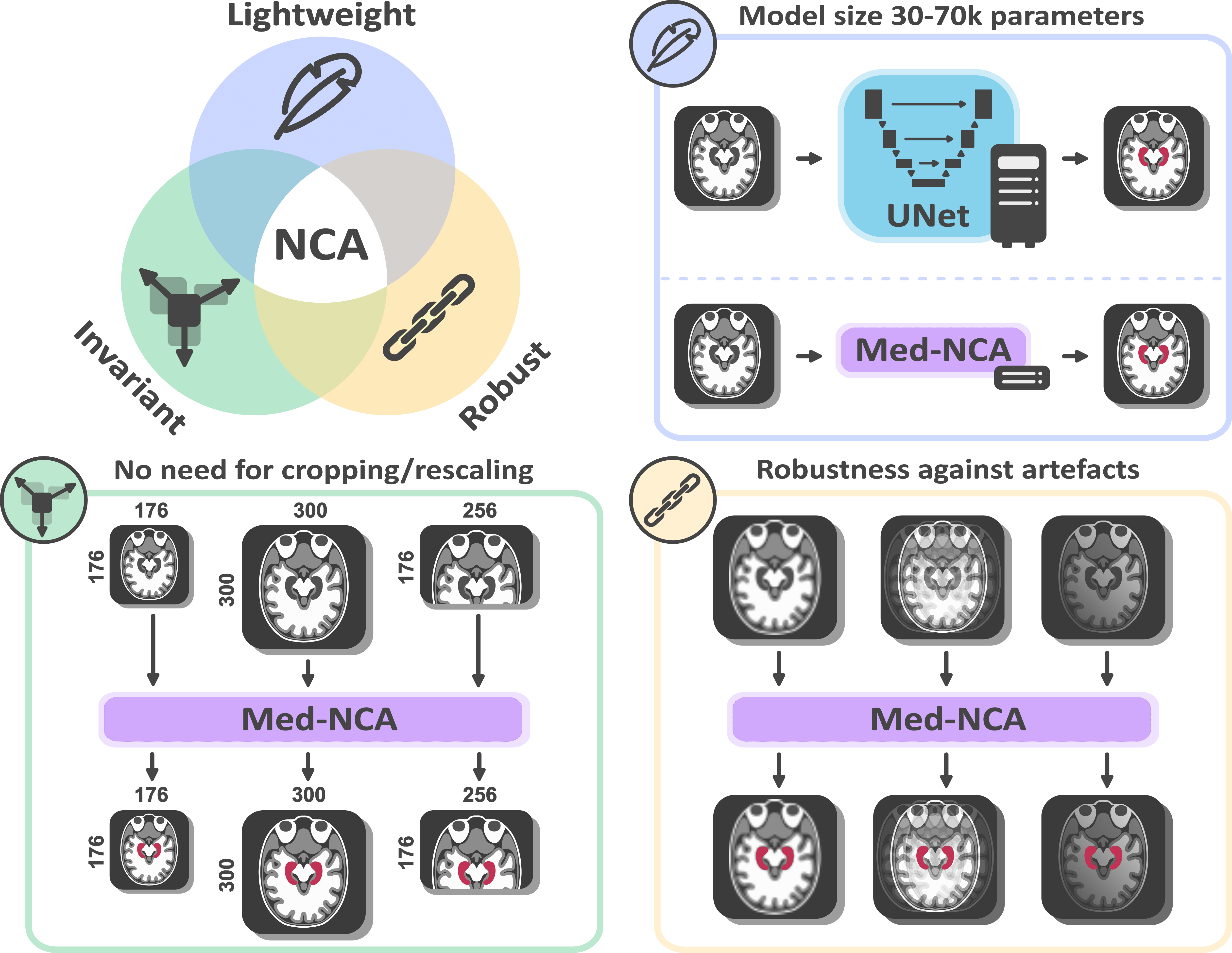}}
  \caption{NCA models are \emph{lightweight} due to their small size and asynchronous inference, and can be run on low-powered systems. They are also, by design, \emph{invariant} to the input scale and field of view. Further, they are \emph{robust}  against image artefacts.}
  \label{fig:general_idea}
\end{figure}

Unlike most state-of-the-art methods that rely on optimising UNet-based frameworks, such as the nnUNet pipeline \cite{nnUnet}, we investigate a fundamentally different learning system that is by design \textbf{lightweight}, \textbf{robust} and \textbf{input-invariant}, yet achieves \textbf{reliable performance}. We introduce Neural Cellular Automata (NCA) as our base architecture, which due to the \emph{one-cell model size}, can be \textit{distributed across any image size} during deployment. The minimal size and asynchronous inference \textit{requires significantly less computing power} than classical models. Additionally, due in part to the limited amount of parameters, a rule has to be learnt that \textit{generalises well} across the problem space, which renders it \textit{robust} by design (illustrated in Figure \ref{fig:general_idea}).

Heavily inspired by the interaction between cells in living organisms, NCAs are minimal models that look at a single cell at a time and can only communicate with their direct neighbours. Global knowledge can be transmitted by deploying the model on each cell and iteratively applying the same rules. Each iteration increases the perceptive range by one cell in each direction. Recently, NCAs have made advances in tasks like robust image generation \cite{growingNCA} and even segmentation for small-resolution natural images \cite{segmentationGoogleNCA}, all while learning a single local update rule that is applied incrementally to each cell.


Despite its advantages such as lightweight inference, training NCAs requires \textit{exponentially higher video ram (VRAM)} depending on the input size during training, which quickly reaches 20 GB for a single sample with a resolution of $256 \times 256$. In medical image processing, this is prohibiting as data is typically high in resolution. The local interaction makes inference on big images difficult (greater than $100 \times 100$), as many steps are required to communicate global knowledge. In addition, \textit{high-resolution images increase the convergence difficulty}. Due to these constraints, previous works on NCAs have focused on small-resolution computer vision benchmarks \cite{medicalsegmentationdecathlon,hernandez2021neural,growingNCA,niklasson2021self,palm2021variational,randazzo2020self}. We solve these limitations with \textit{Med-NCA}, a two-step NCA model. The model distributes global knowledge across a downscaled image in the first step. In the second step, Med-NCA combines the resulting information with high-resolution image patches to perform high-quality segmentations.

We evaluate our proposed Med-NCA on T1-weighted hippocampus and T2-weighted prostate MRI datasets. We first compare the segmentation performance of Med-NCA to classic and efficient UNet-style architectures, where Med-NCA outperforms them by at least 2\% for the hippocampus and 3\% for the prostate, with a \textit{90 to 500 times smaller model size}, although there is still a 2\% and 10\% performance gap to the auto ML pipeline nnUNet. Secondly, we perform an in-depth analysis of three types of \textbf{input invariances}: scale, shape and translation. Med-NCA shows \textit{consistent performance} in comparison to UNet-style models and can even outperform the nnUNet for strong shifts in shape and translation. We then investigate the influence of synthetic MRI acquisition artefacts of increasing severity on Med-NCA and UNet. Despite the vastly different one-cell local interaction setup of Med-NCA, our experiments show similar \textbf{robustness} to the UNet in terms of anisotropy and bias field and even slightly better robustness to ghosting artefacts. Lastly, we demonstrate that \textbf{deployment in low-resource environments} is possible, due to the asynchronous inference and the one-cell model size of NCAs, on a Raspberry Pi Model B\Plus 
 (US\$ 35). There is currently a slightly stronger successor, the Raspberry Pi Zero, which costs US\$ 5.




To ensure reproducibility and drive further research on NCA segmentation, \textit{we make our complete framework available} under \url{github.com/MECLabTUDA/Med-NCA}.

\section{Related Work}
Recent publications have shown the applicability of NCAs to different tasks, such as robust image generation from a single cell \cite{growingNCA} and foreground-background segmentation \cite{segmentationGoogleNCA}. In this section, we review relevant related work on NCAs and medical image segmentation. \\
\textbf{Neural Cellular Automata: }
NCA models are a one-cell model architecture, recently adapted to convolutional neural networks by Gilpin \cite{gilpin2019cellular}. NCAs do not look at the whole image globally but instead only interact locally. Each cell can exclusively communicate with its direct neighbours, and all inherit the same learnt rule. By performing multiple iterations, global knowledge can be conveyed between cells. Despite their small size, NCAs have shown robustness in tasks such as image generation \cite{growingNCA,palm2021variational}, where models display a high degree of resilience against perturbations. To the best of our knowledge, only one previous work explores image segmentation with NCAs \cite{segmentationGoogleNCA}. The proposed method focuses on foreground-background segmentation on small images of 64x64. While it provides a simple up/downscaling solution for high-resolution images, performance is insufficient for medical image segmentation (see Table \ref{tab:AblationStudy}). \\
\textbf{Medical Image Segmentation: } 
With the improvements in graphics cards and VRAM availability \cite{nnUnet}, machine learning models are growing significantly in size. Models like the state-of-the-art nnUNet define 4GB of VRAM as their minimum requirement \cite{nnUnet}, and thus require proper infrastructure for inference. There have been several attempts to create minimal segmentation models, mainly by modifying the well-established UNet \cite{UNet}. The 'Segmentation Models' python package \cite{segmentation_models_pytorch} provides a collection of UNet-style models where the encoder has been replaced with smaller architectures like \textit{EfficientNet} \cite{efficientnet}, \textit{MobileNetV2} \cite{mobilenetv2}, \textit{DenseNet} \cite{densenet}, \textit{ResNet18} \cite{resnet} and \textit{VGG11} \cite{veryDeepConvolutionalNetwork}. However, computational requirements are still significant (see Figure \ref{fig:NCA_DiceVsParameters}). 

UNets and other state-of-the-art segmentation models typically have a pyramid-like structure with multiple up- and downscaling blocks. NCAs stand in strong contrast as they are tiny models acting on a single pixel and communicating global knowledge through iteratively applying the same rule. This intrinsic change in the design allows NCAs to maintain a number of parameters several orders of magnitude lower, and consequently to be run on minimal hardware.

\section{Methodology}
The local architecture of NCAs, where the model deals only with a single cell and its surroundings, allows them to be \emph{lightweight} in terms of storage space and inference time, but this does not come without limitations. Training NCAs end-to-end on images of size 256x256 can easily require 20GB of VRAM (for batch size 1). This is because NCAs are replicated across all input image pixels, and backpropagation is performed through all the iterative steps of the model, increasing VRAM needs. VRAM requirements are therefore dependent on model size, input size and number of iterations/steps.


An additional consideration is that increasing the difficulty of the problem (e.g. directly learning on the full-scale high-resolution image) \emph{can result in NCAs not converging}. Med-NCA reduces VRAM needs and simplifies the segmentation problem by separating it into two steps (illustrated in Figure \ref{fig:highres_training}). Its standard configuration reduces the required VRAM for training by a factor of 16 compared to a full-resolution learning setup and enough steps to increase the perceptive range of each cell to a global scale. This could be improved even further by adding one or more downsampling steps into the pipeline if the need arose.


\subsection{Med-NCA}
\begin{figure}[htbp]
  {\includegraphics[width=1.\linewidth]{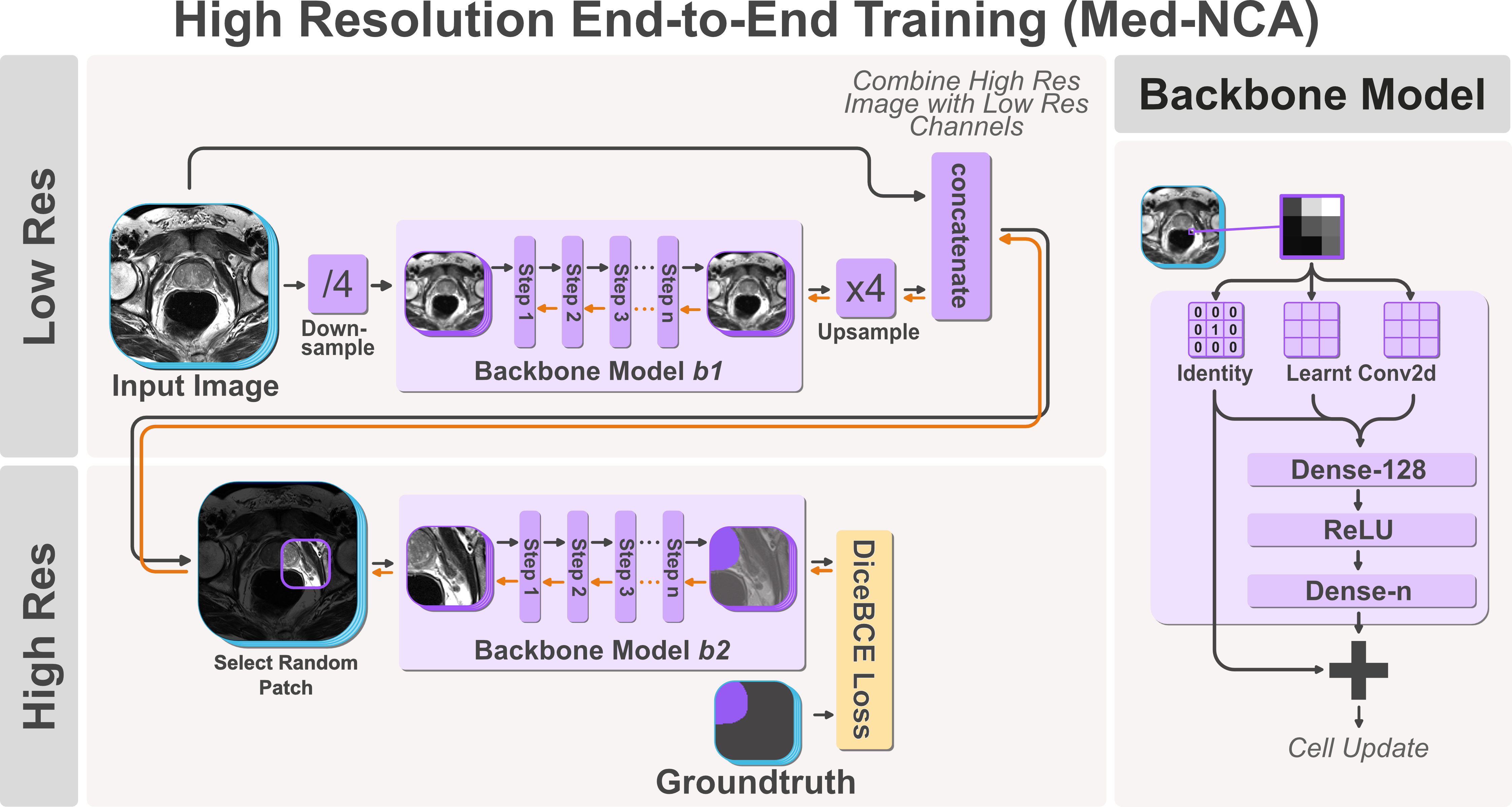}}
  \caption{The \emph{Med-NCA} training strategy relies on cropped images in the 'High Res' step of segmentation to limit VRAM requirements. The final inference is performed on the full image and does not require patchification.}
  \label{fig:highres_training}
\end{figure}

Med-NCA is our main methodological contribution defining a pipeline for training NCAs on high-resolution images. It is optimised to reduce VRAM as well as simplify training for bigger images. We illustrate the training procedure in Figure \ref{fig:highres_training}. We start by describing the backbone NCAs we use in Med-NCA, and then describe the training and inference processes.

\subsection{Backbone NCA}
Med-NCA consists of two identical backbone NCA models that iterate over different scales of the input image. Our models are inspired by the architecture presented in \textit{Growing Neural Cellular Automata} \cite{growingNCA}. 


The backbone model is constructed of $n$ input channels, where the first $x$ channels are reserved for the image. The NCA can freely set the remaining $n$ - $x$ channels. Instead of growing from a single pixel, we adapt it to the segmentation task by immediately distributing it across the whole image. 

The proposed model consists of two 3x3 learnt convolutional layers and is concatenated with the current cell state, resulting in a state vector of size $3*n$, as illustrated in Figure \ref{fig:highres_training}. It is important that the learnt convolutional layers use reflect padding, as the model otherwise learns to use the image borders for 'spatial orientation', which worsens input invariance capabilities. The state vector is connected to a dense layer with hidden size $h$, following a ReLU and another dense layer with the output size $n$. The standard configuration of our model uses the following parameters: $n = 32$, $h = 128$. These parameters are set to the maximum value at which the model still converges stably, as this provides cells with more memory and allows them to learn a more advanced update rule. For a version of the model that requires 2.5 times fewer parameters, the channel size can be set to $n=16$, resulting in a slight performance degradation, as shown in the ablation study in Table \ref{tab:AblationStudy}. Similar to other NCA approaches \cite{growingNCA,segmentationGoogleNCA}, we randomly activate $x=50\%$ of the cells in each step to simulate asynchronous activation.


\subsection{Training of Med-NCA}
Med-NCA uses multiple-level NCAs for performing the segmentation. We do that by training two backbone NCAs $b1$ and $b2$ on different image scales so that training remains stable and the model learns to consider global and detail-rich information.


The pipeline is executed as follows: in the first step, the image is downsampled by a factor of four. Then, $b1$ is iteratively applied for $s$ steps on the input image $x$. Afterwards, the output $\bar{x}$ is upscaled back to the original image size. In the next step, we replace the first channel of the output, still containing the upscaled low-resolution image, with its high-resolution counterpart, thus allowing $b2$ to refine the outputs further. We then take a random patch $p$ that is of similar size as $x$ and iteratively apply $b2$ $s$ times. Lastly, we perform backpropagation with a loss of Dice and binary cross-entropy on the patch prediction and the corresponding ground truth segmentation.

This two-step process lowers VRAM requirements by a factor of 16 during training, making it possible to train Med-NCA on high-resolution images.

\subsection{Inference}
While training has to be specifically adapted to work well with big images, inference is very simple. Extracting a prediction is extremely lightweight and can be carried out on nearly any device that runs PyTorch. The asynchronous and iterative nature of NCAs makes this possible. Inference is only limited by the RAM size, where Med-NCA has a much smaller footprint than, e.g., a UNet. Since NCAs itself are very small, only the size of the image and number of channels are relevant. Further, \textit{during inference patchification is not necessary} as the trained model can be applied directly to the whole image.

\section{Experimental Setup}
The evaluation of our Med-NCA focuses on the three main aspects \textit{robustness}, \textit{input invariance} and \textit{model size}. The evaluation is performed on two segmentation tasks, namely hippocampus and prostate. We compare our proposed method with UNet-style architectures as well as the auto ML pipeline nnUNet.

\textbf{Data:} We use the hippocampus data released for the Medical Segmentation Decathlon (MSD) \cite{medicalsegmentationdecathlon}. The prostate data is a mixture of two datasets, the MSD (CC-BY-SA 4.0 licence) and ISBI 2013 challenge (CC BY 3.0) \cite{nci_isbi_challenge}. The prostate images range from $320 \times 320$ to $384 \times 384$. We scale the image to a training size of $256 \times 256$, allowing us to perform out-of-distribution experiments for smaller and higher-resolution scales. \\
\textbf{Evaluated architectures:} We compare the performance of our proposed Med-NCA to the well-established UNet \cite{UNet} and different resource-efficient versions. In the latter case, the encoder is replaced by \textit{EfficientNet} \cite{efficientnet}, \textit{MobileNetV2} \cite{mobilenetv2}, \textit{DenseNet} \cite{densenet}, \textit{ResNet18} \cite{resnet} and \textit{VGG11} \cite{veryDeepConvolutionalNetwork}. We use the version of each encoder with the least amount of parameters provided in the \textit{Segmentation Models} repository \cite{segmentation_models_pytorch}. We also compare our approach to 2D and 3D full-resolution versions of the nnUNet \cite{nnUnet}.

\section{Results}
The evaluation of our proposed Med-NCA model shows that it deals well with MRI acquisition artefacts and input size variations while reaching state-of-the-art performance on the hippocampus and prostate segmentation datasets. \\

\subsection{Performance and Resource Consumption}
\begin{figure}[htbp]
  \centering
  \includegraphics[width=0.9\linewidth]{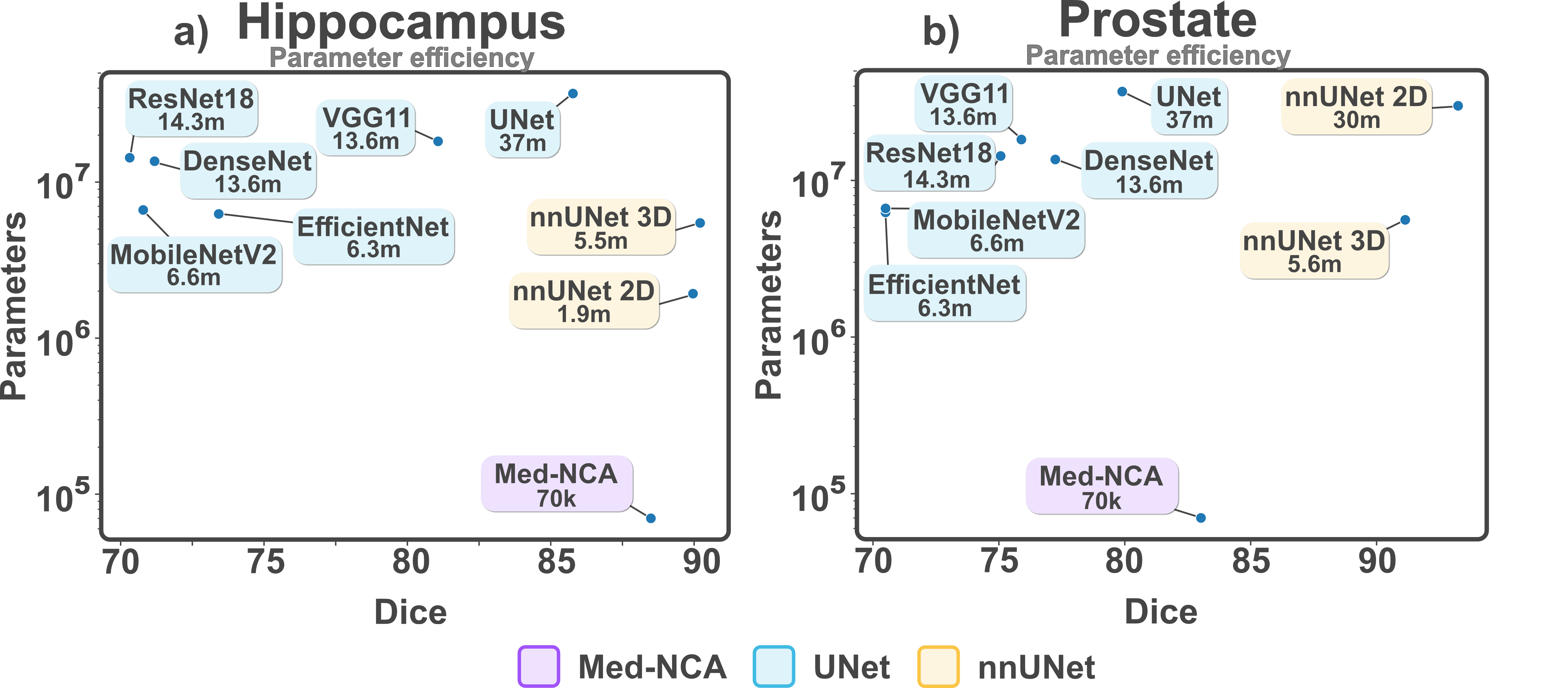}
  \caption{We compare Med-NCA to other efficient UNet setups as well as the nnUNet in terms of performance vs. the number of parameters.}
  \label{fig:NCA_DiceVsParameters}
\end{figure}

We perform a thorough performance analysis of Med-NCA compared to other UNet-style architectures, and place it in the context to resource consumption. This relation is illustrated in Figure \ref{fig:NCA_DiceVsParameters}. 

\begin{wrapfigure}{r}{0.5\textwidth}
\centering
\includegraphics[width=0.8\linewidth]{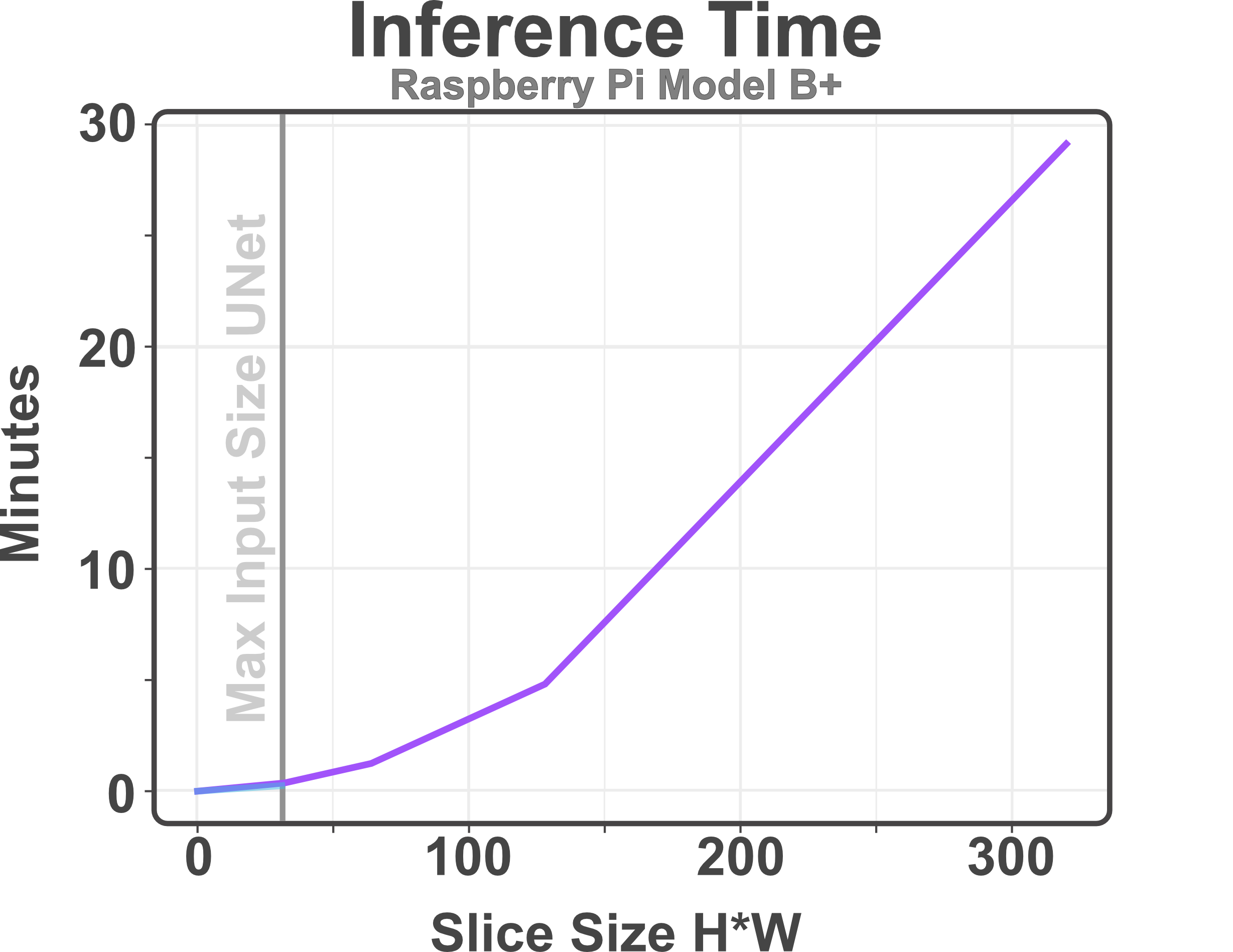} 
\caption{Med-NCA single slice segmentation time on a Raspberry Pi B\Plus by image size.}
\label{fig:wrapfig}
\end{wrapfigure}
We see a general trend of UNet-style models suffering in performance from a decrease in model size. Med-NCA, on the other hand, reaches higher Dice scores than the classic UNet, outperforming it by 2\% on the hippocampus data and 3\% on the prostate dataset, while also requiring 500 times less trainable parameters. 

To show how lightweight Med-NCA is we deploy it on a Raspberry Pi B\Plus and perform inference up to an image size of $320 \times 320$ (see Figure \ref{fig:wrapfig}). While inference is rather slow, with 30 min per slice for the maximal image size of $320 \times 320$, using a more recent system with an \textit{RTX 2060 Super} and an \textit{AMD Ryzen 5600X} inference only takes seconds for a whole MRI. In comparison, the UNet can only be executed on the Raspberry Pi up to an image size of $32 \times 32$.

\subsection{Input Invariance} 
\begin{figure}[htbp]
  {\includegraphics[width=1\linewidth]{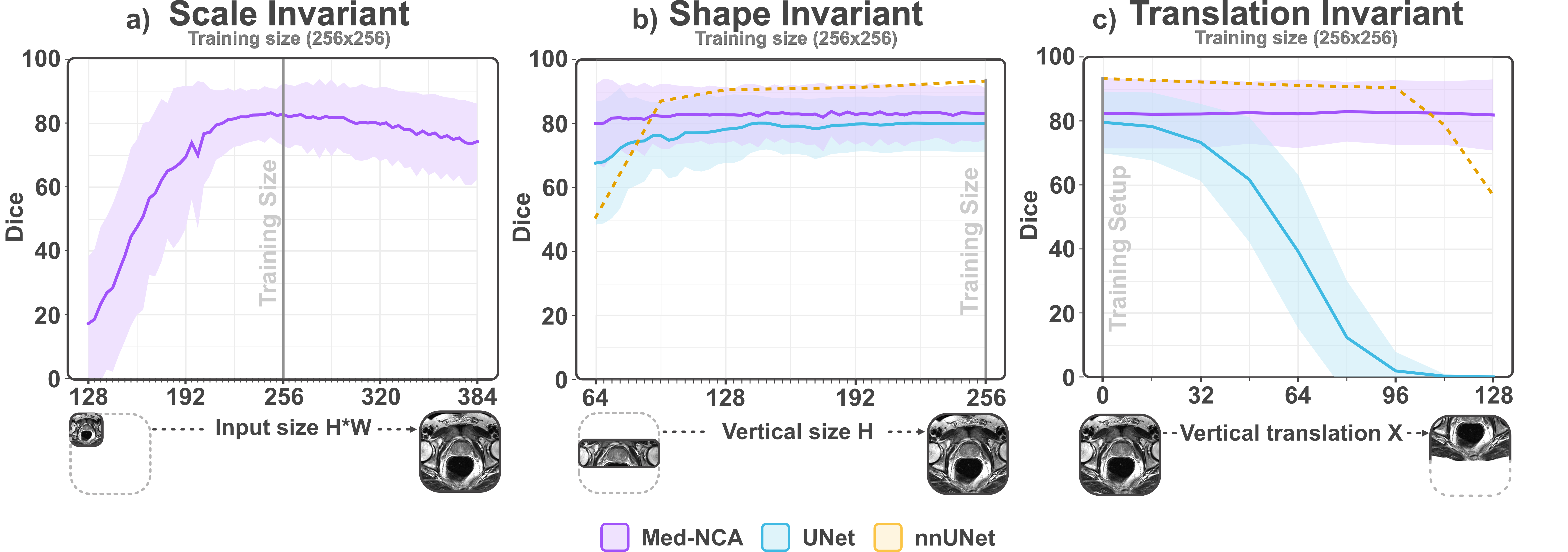}}
  \caption{Comparison of our proposed Med-NCA with UNet and the nnUNet in of out-of-distribution (a) scale, (b) shape and (c) translation scenarios.
  }
  \label{fig:input_invariance}
\end{figure}

When dealing with medical images, the scale or input size may change after training. A problem with UNet-style segmentation models is that they are not good at adapting to such variability, which our experiments in Figure \ref{fig:input_invariance} show. When UNet faces strong shifts in vertical shape, the Dice drops by 13\%, whereas Med-NCA only loses 3\%. While the auto ML pipeline nnUNet performs robustly until strong severity of shape changes, it then loses even more performance and drops to 50\% Dice. We experience similar results for translational changes. While Med-NCA performs consistently across all introduced shifts, the UNet drops to 0\% Dice, indicating strong overfitting on the position of the image. As for shape variations, the nnUNet performs stable until strong translational shifts appear, where it again drops to 56\% Dice. The third experiment we conduct is Med-NCAs capability of dealing with different image sizes as it can be arbitrarily distributed across any image size due to its one-cell model setup. Med-NCA shows only slight performance losses of maximal 3\% when dealing with images that are in the range of 0.8 up to 1.5 times the training size.
When working on MRIs, input invariance is especially relevant as only scanning a partial region or lowering the resolution can \textit{greatly increase capturing speed}. Med-NCA has the considerable advantage that it can be trained on a full upper body scan but perform equally on partial images. \\

\subsection{Robustness} 
\begin{figure}[htbp]
  {\includegraphics[width=1.\linewidth]{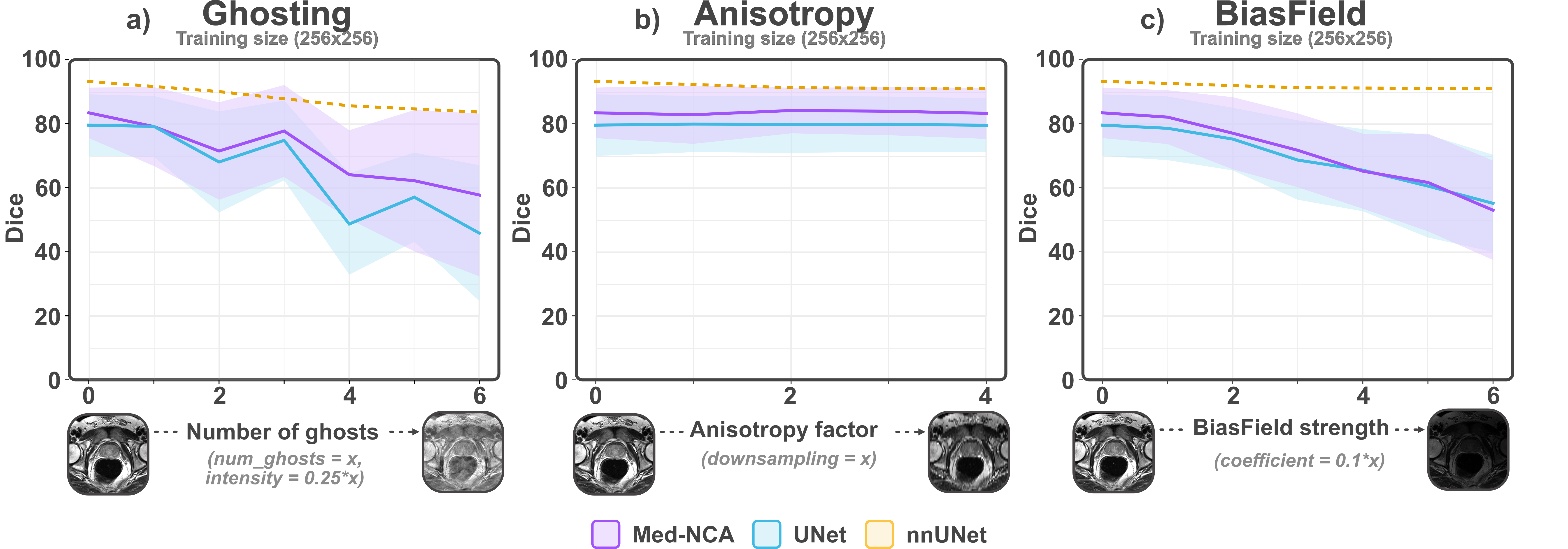}}
  \caption{Robustness analysis of our proposed Med-NCA in comparison to UNet and the nnUNet. nnUNet results are collected for fewer data points, indicated by the dashed line, as the non-flexible pipeline requires manual evaluation. Analysis is performed for synthetic MRI acquisition artefacts of increasing severity: (a) Ghosting, (b) Anisotropy and (c) Bias Field (using TorchIO \cite{perez-garcia_torchio_2021}).}
  \label{ImageArtefacts}
\end{figure}

Robustness against MRI acquisition artefacts is an important trait in medical image segmentation. In our robustness analysis, illustrated in Figure \ref{ImageArtefacts}, we show that Med-NCA can handle anisotropy and bias field artefacts similarly well to a classic UNet. Both models experience no drop in performance with anisotropy artefacts up to a severeness factor of 4. In the case of bias field, Med-NCA and UNet drop more than 25\% in performance when the severity of the artefacts becomes too strong. In cases of severe ghosting artefacts, we can see the performance of Med-NCA suffers less than a classic UNet, where Med-NCA performs 12\% better in the most severe cases of ghosting. 

As our comparison to the nnUNet demonstrates, an auto ML pipeline can increase the robustness as nnUNet only suffers from slight performance losses across the MRI acquisition artefacts. It is plausible to include a similar setup in the training of Med-CNA and thus improve robustness.

\begin{table}[htbp]
\begin{center}
\begin{tabular}{ l |c |c | c | c }
 \textbf{Model} & \multicolumn{2}{c|}{\textbf{Hippocampus}} & \multicolumn{2}{c}{\textbf{Prostate}} \\

 & \textbf{Dice} $\uparrow$ & \textbf{\# Param.} $\downarrow$ & \textbf{Dice} $\uparrow$ & \textbf{\# Param.} $\downarrow$ \\
 \hline
 \hline
 \hline
 Med-NCA & \textbf{$\mathbf{\numprint{0.885514788}\pm\numprint{0.04185757813}}$} & 70016 & \textbf{$\mathbf{\numprint{0.8381750815}\pm\numprint{0.08262496859}}$} & 70016 \\
 Med-NCA c = 16 & $\numprint{0.8728768748748621787}\pm\numprint{0.03318942062}$ & \textbf{25920} & $\numprint{0.8224211633}\pm\numprint{0.08680814187}$ & \textbf{25920} \\
 Med-NCA h = 64 & $\numprint{0.8577370057182927}\pm\numprint{0.04681903125}$ & 47530 & $\numprint{0.8077421718}\pm\numprint{0.1252152891}$ & 47530 \\
 Backbone-NCA 64x64 & $\numprint{0.8710985875898792}\pm\numprint{0.870272526}$ & 35008 & $\numprint{0.7890843484136794}\pm\numprint{0.131686187}$ & 35008 \\
 Seg. NCA \cite{segmentationGoogleNCA} & $\numprint{0.8049988622}\pm\numprint{0.0454523702}$ & 39472 & $\numprint{0.6338137057}\pm\numprint{0.1900379025}$ & 39472\\
  UNet & $\numprint{0.8577370057182927}\pm\numprint{0.04443966322}$ & 36951555 & $\numprint{0.7989924848079681}\pm\numprint{0.0989710729}$ & 36951555\\ 
 \hline
\end{tabular}
\newline
\caption{Comparison of different Med-NCA setups (where $c = channels$, $h = hidden size$), as well as the Backbone-NCA, previous work on NCA segmentation and a standard 2D UNet.}
\label{tab:AblationStudy}
\end{center}
\end{table}

\subsection{Ablation Study} Lastly, we perform an ablation on our approach in Table \ref{tab:AblationStudy}. Our results show that a different trade-off of performance vs. model size can make Med-NCA 2.5x more lightweight while only sacrificing 1\% of performance. Further, we can see that the previous NCA segmentation model \textit{Seg. NCA} \cite{segmentationGoogleNCA} is not suitable for the task of medical image segmentation as it performs 7.4\% worse for hippocampus and even 20.2\% worse for the prostate segmentation task. \\

\section{Discussion}
We have shown that NCA-based architectures are not only suitable for low-resolution imaging tasks, but can also be leveraged for high-resolution image segmentation with our proposed Med-NCA. Since standard NCAs require exponential amounts of VRAM, determined by the input size, this requires an adapted training pipeline. We enable the training with high-resolution images by incorporating patches in the second segmentation step during training. Inference can be performed directly on the full-resolution image. 

When comparing Med-NCA to the classical UNet and resource-efficient variations, Med-NCA outperforms them by 2\% and 3\% Dice for hippocampus and prostate image segmentation, respectively. Med-NCA is also more robust to changes in image scale, shape and translation, which is directly inherited by its one-cell model size. The local interaction prevents the model from knowing where in space information is located and therefore eliminates biases based on position. Further, despite the vastly different one-cell model size with 500 times fewer parameters, Med-NCA shows similar or better robustness than a UNet to MRI acquisition artefacts. Due to the one-cell model size and asynchronous inference, NCAs are lightweight enough to be executed on any hardware that runs PyTorch, which we demonstrate by deploying Med-NCA on a Raspberry Pi B\Plus, that can be powered by any 5 Watt power source like a standard power bank. 

While we have solved the problem of NCAs for high-resolution inputs, one limitation of the present approach is that it only admits 2D image slices. Further optimisations could make Med-NCA work for 3D inputs by adapting the perceptive field and integrating VRAM improvements based on the three-dimensional space. 3D inputs would make Med-NCA suitable for more datasets and might further improve performance. In addition, although Med-NCA is inherently robust, we experience performance degradation when the severity of the acquisition artefacts becomes too strong, which could be improved upon in future work. 

A further limitation is that there is a 10\% performance difference between Med-NCA and nnUNet on the prostate task. For future work, we propose the development of a training pipeline similar to the nnUNet, which uses NCA as the core architecture and includes data augmentations and post-processing steps during training. We expect that such a framework could achieve equivalent performance while profiting from the inherited benefits of the NCA architecture like size and scale invariance and general robustness. \\

\section{Conclusion}

In this work we introduce the Med-NCA segmentation model, which solves the VRAM limitations inherited by NCAs for high-resolution images and is therefore suitable for medical images. Our approach can be run on minimal hardware (e.g. a Raspberry Pi B\Plus), is inherently robust against scale, shape, translation and image artefacts, and reaches near state-of-the-art segmentation performance. We compare our model to UNet-style models optimised for low parameter size and show that Med-NCA not only achieves a higher Dice score, but also outperforms them in terms of input invariance, robustness and resource use. This makes Med-NCA a perfect candidate for primary care scenarios with limited infrastructure and highly variable imaging equipment.





\bibliographystyle{splncs04}
\bibliography{bibliography}

\end{document}